\documentclass[showpacs,preprintnumbers,amsmath,amssymb,twocolumn]{revtex4}
\usepackage{graphicx}
\usepackage{makeidx}
\usepackage{chemarr}
\usepackage{bm}\newcommand{\C}{{\cal{C}}}
\newcommand{\Cp}{{\cal{C^{\prime}}}}
\def\prl#1#2#3{{ Phys. Rev. Lett.} {\bf #1}, #2 (#3)}
\def\prep#1#2#3{{ Phys. Rep.} {\bf #1}, #2 (#3)}

\def\pre#1#2#3{Phys. Rev. E {\bf #1}, #2 (#3)}

\def\pnas#1#2#3{Proc. Natl. Acad. Sci. (USA) {\bf #1}, #2 (#3)}

\def\jpa#1#2#3{J. Phys. A {\bf #1}, #2 (#3)}
\def\jcp#1#2#3{J. Chem. Phys. {\bf #1}, #2 (#3)}
\def\jpc#1#2#3{J. Phys. Chem. {\bf #1}, #2 (#3)}

\def\physa#1#2#3{Physica A {\bf #1}, #2 (#3)}

\def\tree#1#2#3{Trends Ecol. Evol. {\bf #1}, #2 (#3)}
\def\ch#1#2#3{Chaos {\bf #1}, #2 (#3)}
\def\nat#1#2#3{Nature {\bf #1}, #2 (#3)}
\def\mol{\mbox{mol}}
\def\hr{\mbox{h}}
\def\p{^{\prime}}
\def\h{\mbox{h}}
\def\DA{D_{A}}
\def\DAP{D_{A}^{\prime}}
\def\DR{D_{R}}
\def\DRP{D_{R}^{\prime}}
\def\DS{D_{S}}
\def\DSP{D_{S}^{\prime}}
\def\MA{M_{A}}
\def\MR{M_{R}}
\def\MS{M_{S}}
\def\ta{\theta_{A}}
\def\tr{\theta_{R}}
\def\ts{\theta_{S}}
\def\ga{\gamma_{A}}
\def\gr{\gamma_{R}}
\def\gs{\gamma_{S}}
\def\aa{\alpha_{A}}
\def\apa{\alpha_{A}^{\prime}}
\def\ar{\alpha_{R}}
\def\apr{\alpha_{R}^{\prime}}
\def\as{\alpha_{S}}
\def\aps{\alpha_{S}^{\prime}}
\def\ba{\beta_{A}}
\def\br{\beta_{R}}
\def\bs{\beta_{S}}
\def\dma{\delta_{MA}}
\def\dmr{\delta_{MR}}
\def\dms{\delta_{MS}}
\def\da{\delta_{A}}
\def\dr{\delta_{R}}
\def\ds{\delta_{S}}
\def\gc{\gamma_{C}}
\def\gcp{\gamma_{C}^{\prime}}

\def\epsilon{\varepsilon}

\def\x{{\bf{x}}}
\def\f{{\bf{f}}}

\def\beqr{\begin{eqnarray}}
\def\eqnr{\end{eqnarray}}
\def\beq{\begin{equation}}
\def\bc{\begin{center}}
\def\ec{\end{center}}
\def\eqn{\end{equation}}
\begin{document}
\title{The synchronization of stochastic oscillators}
\author{Amitabha Nandi$^1$, Santhosh G.$^1$, R. K. Brojen Singh$^2$,
and Ram Ramaswamy$^{1,2}$}
\affiliation{$^1$School of Physical Sciences, Jawaharlal Nehru University,
New Delhi 110067, India\\
$^2$Center for Computational Biology and Bioinformatics, School of
Information Technology,
Jawaharlal Nehru University, New Delhi 110067, India}
\date{\today}
\begin{abstract}
We examine microscopic mechanisms for coupling stochastic oscillators so
that they display similar and correlated temporal variations. Unlike 
oscillatory motion in deterministic dynamical systems, complete 
synchronization of stochastic oscillators does not occur, but appropriately 
defined oscillator phase variables coincide. This is illustrated in model 
chemical systems and genetic  networks that produce oscillations in the 
dynamical variables, and we show that  suitable coupling of different 
networks can result in their {\it phase} synchronization.
\end{abstract}
\maketitle
The {concerted} behaviour  of different stochastic processes can be of
considerable importance in a variety of situations. Examples from a
biological context \cite{glass} range from the synchronized firing of groups 
of neurons \cite{neuron} to the occurence of circadian or ultradian 
cycles \cite{church,tu} in groups of metabolic and cellular regulatory 
processes, each of which is individually stochastic. Similar phenomena also 
occur in other areas of study, such as weather modeling or the study of 
coupled populations \cite{lloyd}. It is therefore a moot question whether 
there is a sense in which two or more independent (or unrelated) stochastic 
phenomena can become temporally synchronized. In this Letter we consider 
mechanisms through which such synchronization can be effected, and examine 
measures through which the synchronization can be detected.

The synchronization of coupled nonlinear oscillators has been studied 
extensively in the past two decades. Both chaotic and nonchaotic oscillator 
systems can show complete synchronization, when all variables of the systems 
coincide, as well as more general forms such as phase, generalized, or lag 
synchronization \cite{kurths}. Studies have examined a variety of 
different coupling schemes and topologies, and the robustness of 
synchronization to noise. In the typical cases examined, the noise 
is external and thus appears largely as additional stochastic terms in 
otherwise deterministic dynamical equations \cite{a,hanggi}.

For the systems we study here the evolution is {\underline {intrinsically}} 
stochastic. Such systems do not follow deterministic equations of motion: 
a master equation  describes the evolution of configurational 
probabilities \cite{master},
\beq
\label{me}
\frac{d}{dt}P(\C,t)  = -\sum_{\Cp} P(\C,t) W_{\C \to\Cp} + \sum_{\Cp}  P(\Cp,t) W_{\Cp \to\C}
\eqn
where in standard notation \cite{osw}, $P(\C,t)$ is the probability of 
configuration $\C$ and the $W$'s are transition probabilities. The 
configurations that are realized as a function of time give a probabilistic 
description of the system as it traverses the state space of the problem.

Consider now two such independent systems, each of which can be described by 
a (reduced) master equation of the above form.  Starting with similar 
configurations, $\C$, the subsequent evolution of each of the subsystems 
will typically be quite distinct. The concern here is to (a) examine 
conditions under which the two subsystems synchronize, namely follow very 
similar paths in the state space, and to (b) detect this phenomenon.

Our main result can be stated in general terms as follows. When the two 
systems are coupled by employing a mediating process, then variables of the 
two subsystems {\it phase--synchronize}, namely they vary  in unison. Such 
synchronization depends on the nature and strength of the coupling and can 
persist even when fluctuations are large.

It is simplest to illustrate this through an example. The stochastic dynamics 
of the Brusselator model, which has been studied extensively \cite{brussel} 
derives from consideration of the following ``chemical'' reactions 
\cite{gillespie}

\beqr
\label{cb}
A_1 &\xrightarrow{c_1}& X \\
A_2 + X &\xrightarrow{c_2}& Y + A_3 \\
2X + Y &\xrightarrow{2c_3}& 3X \\
X &\xrightarrow{c_4}& A_4.
\eqnr

A master equation formalism is exact for this system in the gas phase and in 
thermal equilibrium \cite{megillespie}, and indeed is necessary when 
the number of molecules is small. On the other hand, in 
the thermodynamic limit, when fluctuations are negligible, one can derive a 
set of deterministic reaction--rate equations  for the concentrations of 
species $X$ and $Y$,
\beqr
\label{deter1}
\dot x &=& c_1 - c_2 x + c_3  x^2y -c_4 x \equiv f_x(x,y)\\
\dot y &=& c_2 x -  c_3  x^2y \equiv f_y(x,y).
\label{deter2}
\eqnr
These rate equations usually have to be integrated numerically to determine
the orbits (for the parameters here, a limit cycle). Similarly, the 
corresponding master equation cannot be solved analytically and recourse must be made to  Monte Carlo simulations
\cite{gillespie}. The number of molecules of species $X$ or $Y$
(taking the volume to be unity) will vary in a stochastic manner, giving a
(noisy)  limit--cycle solution to Eqs.~(2-5).

Another Brusselator system (denoted by similar equations, but with primed variables
and parameters, say) would also naturally give a noisy limit cycle solution,
the characteristics of which would depend on the parameters of the problem.
Thus if the parameters $\{c_i\}$ and $\{c^{\prime}_i\}$ are different
the evolution of the two subsystems
will be independent and uncorrelated.

We consider two scenarios where the subsystems are coupled at the microscopic
level. In the ``direct'' case, we take the species $X$ and $X^{\prime}$ to be identical:
the subsystems share a common drive. We find that for any finite volume this
coupling results in the temporal variation of species $Y$ and $Y^{\prime}$ rapidly becoming
highly correlated, even if the fluctuations are large.

Alternately one may consider an ``exchange'' process, when species $X$ and
$X^{\prime}$ can interconvert. This introduces an additional reaction,
\beq
X\xrightleftharpoons[c\p]{c} X\p
\eqn
that serves to couple the subsystems, and depending on the rate of interconversion
(governed by $c$ and ${c^{\prime}}$), species $Y$ and $Y^{\prime}$ show
synchronization.

\begin{figure}
\scalebox{0.5}{\includegraphics{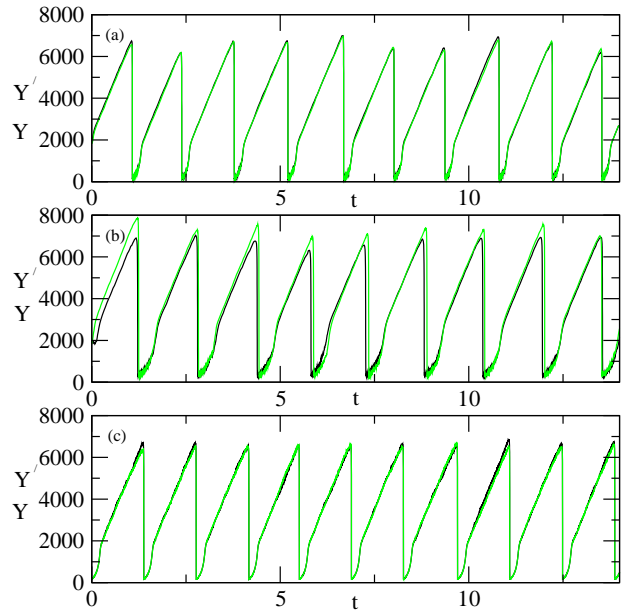}}
\caption{ (Colour online)
Species $Y$ and $Y^{\prime}$ as a function of  time for the coupled
Brusselator system. The  parameter $c_2=50$ differs from
$c^{\prime}_{2}= c_2+5$. The other parameters are $c_1=c^{\prime}_{1}=5000$,
$c_3=c^{\prime}_{3}=0.000025$ and $c_4=c^{\prime}_{4}=5$. (a) Stochastic
simulation results for  $Y$ and $Y^{\prime}$ using direct coupling.
(b) As in (a), but for the case of exchange coupling, Eq.~(8) with
$c=c\p=0.6$. (c) Results obtained by solving the chemical Langevin equation
\cite{lange} with  $\gamma=0.01$.}
\label{fig1}
\end{figure}

Although both direct as well as exchange coupling mechanisms  effect synchronization,
they differ in the details of how they act. Fig.~\ref{fig1} shows the
variation of species  $Y$ and $Y\p$  as a function of  time,  from stochastic
simulations as well as the so--called chemical Langevin approach \cite{lange},
which effectively adds stochastic noise to the
deterministic dynamics, Eqs.~(\ref{deter1}-\ref{deter2}). Clearly the two
concentrations vary in unison. However, due to intrinsic stochastic
fluctuations the two solutions do not become identical but only
phase--synchronize \cite{rpk} as we now show.

To judge the phase synchronization of two stochastic oscillators, it is
necessary to first obtain the phase for a single oscillator. The
Hilbert phase of a signal $s(t)$ is obtained \cite{rpk} by first computing its Hilbert transform,
\beq
\pi\bar s(t) = \mbox{P. V.~}\int_{-\infty}^{\infty}\frac{s(\tau)}{t-\tau}d\tau
\eqn
(P. V. denotes principal value) and defining the instantaneous amplitude $A(t)$ and phase $\phi(t)$ through
\beq
A(t) e^{i\phi(t)} = s(t)+ i \bar s(t).
\eqn

\begin{figure}
\scalebox{0.5}{\includegraphics{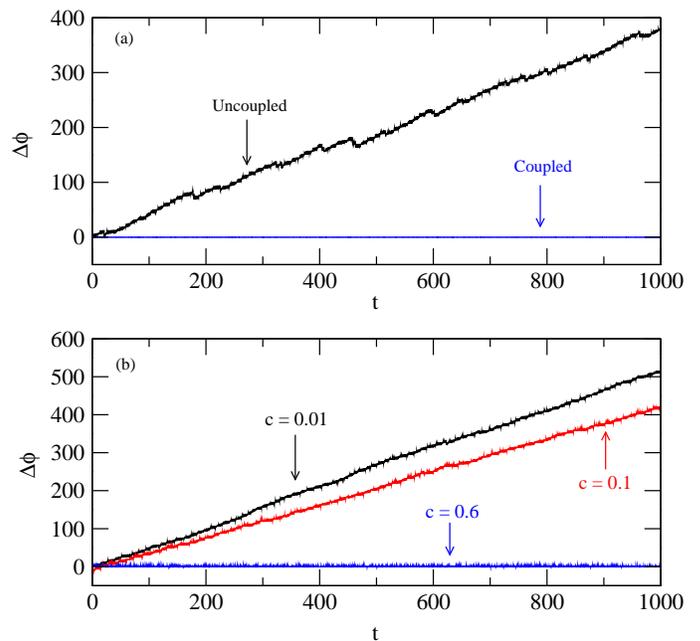}}
\caption{(Colour online)
Phase difference with parameter mismatch for the Brusselator model for
(a) direct coupling of species $X$ and $X^{\prime}$, and
(b) exchange coupling, for different strengths.}
\label{fig2}
\end{figure}

The difference in the Hilbert phases $\phi$ and $\phi^{\prime}$ of the two
stochastic oscillators is shown in Fig.~\ref{fig2}. When the subsystems are
uncoupled, the phase difference $\vert \phi - \phi^{\prime}\vert$ increases
linearly in time on average since the oscillators evolve independent of one
another. With coupling the two oscillators
phase--synchronize and the phase difference fluctuates around  a constant
value.

\begin{figure}
\scalebox{0.25}{\includegraphics{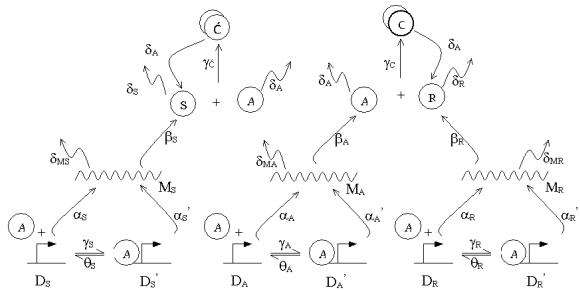}}
\caption{
Biochemical network of the extended circadian oscillator model. $\DA$ and $\DAP$
denote the number of activator genes with and without $A$ bound to its
promoter respectively, and  $\DR$, $\DRP$ and $\DS$, $\DSP$, refer to
the two repressors driven by the common promoter $A$. $\MA$, $\MR$ and $\MS$
denote mRNA corresponding to the activator  $A$, and the repressors $R$ and $S$.
$C$ and $C^{\prime}$ corresponds to the inactivated complexes formed by $A$ and
$R$, and $A$ and $S$ respectively. The constants $\alpha$ and
$\alpha^{\prime}$ denote the basal and activated rates of transcription,
$\beta$ the rates of translation, $\delta$ the rates of spontaneous
degradation, $\gamma$ the rates of binding of $A$ to other components, and
$\theta$ denotes the rates of unbinding of $A$ from those components. 
See \cite{param}.}
\label{bioch}
\end{figure}

\begin{figure}
\scalebox{0.4}{\includegraphics{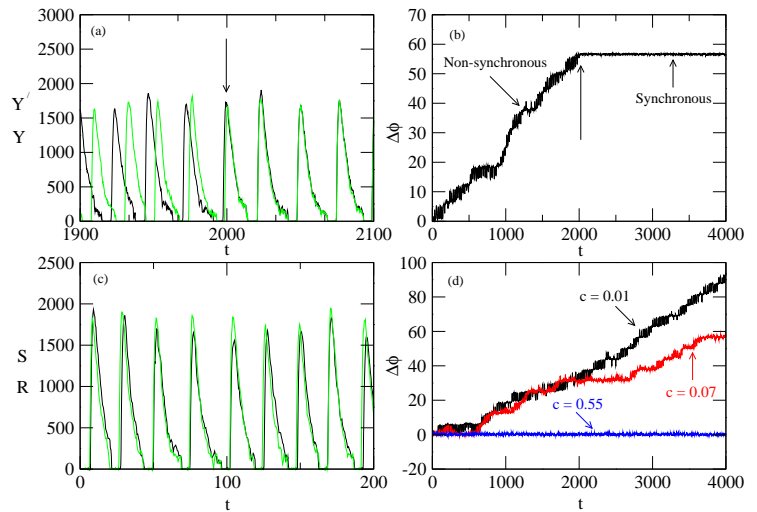}}
\caption{(Colour online)
Temporal behaviour and phase difference of the repressors in the
circadian oscillator model. (a) The repressors are initially  uncoupled
and the coupling is switched on (vertical arrow) at time $t \ge 2000$.
(b) The corresponding phase difference, which
becomes constant for  $t \ge 2000$. (c) Time-series of the  repressors
oscillating in unison for  diffusive coupling with  $c=c\p= 0.55$. (d) Phase
differences for various $c$.}
\label{fig4}
\end{figure}

The coupling schemes discussed here can bring about stochastic
synchronization in a very general setting. Consider, for example, a model
circadian oscillator that has been quantitatively studied in some detail
recently \cite{genosc}.  Shown in
Fig.~\ref{bioch} is the biochemical network for two such oscillators coupled
with a common drive, namely two genetic circuits with a common activator.
In effect a single activator binds to two promoter sites
for repressor proteins $R$ and $S$, a fairly commonplace situation.
Each individual circuit gives stochastic oscillations in the number of
repressor molecules. When the two systems are coupled the  stochastic
oscillations of the two subsystems rapidly phase--synchronize. The
synchronization is robust to parameter variation: in the examples shown in
Fig.~\ref{fig4}, the corresponding parameters of the two subsystems differ by
as much as 10\%; nevertheless the variables of the two systems oscillate in
phase in a stable and sustained manner.

The temporal behaviour of the two repressors and their phase difference for
both direct  and exchange coupling are shown in Fig.~\ref{fig4}. In (a) the
two systems are initially uncoupled and therefore evolve independently. The
coupling is switched on for $t \ge 2000$, leading rapidly to a constant phase
difference, indicative of the phase synchronization (Fig. \ref{fig4}(b)). In the case
of   exchange coupling the genetic circuit differs somewhat from Fig.~\ref{bioch}: the
circuit of Ref. \cite{genosc} is essentially doubled, and there is an
additional activator $A^{\prime}$. Through the equivalent of Eq.~(8)  the activators of
the two circuits are allowed to interconvert, and the two subsystems
synchronize as can be seen in Fig. \ref{fig4}(c-d).

Examination of the macroscopic dynamics offers some clues as to how the
stochastic oscillators synchronize. While the master equation
description is formally exact at the microscopic level, the kinetic equations
that describe the  system at a macroscopic level can be derived from 
Eq.~(\ref{me}) by first obtaining the generating function representation, 
followed by a cumulant expansion \cite{nicolis} to give a set of ordinary 
differential equations,
\beq
\label{nde}
\dot \x  = \f (\x,t)+ O(\frac{1}{V})+ O(\frac{1}{V^2})+ \ldots
\eqn
in the variables $x_i={\langle X_i\rangle}/{V}$,
namely the average concentrations,  with ``noise'' corrections that depend on
the  system volume $V$. Coupling two such systems as discussed above would give a set of
coupled stochastic differential equations, and the circumstances under which these will phase synchronize have been studied to some extent \cite{kurths,neiman}.  
The direct coupling, in the deterministic limit \cite{gillespie,cle} leads 
to dynamical equations which are similar (but not identical) to those 
that derive from the coupling scheme proposed by Pecora and Carroll 
\cite{pecora} for the synchronization of chaotic oscillators.
Although it is not possible to define Lyapunov exponents for stochastic
systems, analysis of the deterministic limit gives indications of what
coupling schemes could be effective. Making species
$Y$ the common drive between the two Brusselator subsystems does not serve
to synchronize them but instead leads to the stochastic analogue of oscillator death.

Similarly, the exchange process, Eq.~(8) results in
diffusive coupling \cite{boccaletti} in the kinetic equations for the species $X$
and $X^{\prime}$, namely terms of the form $c(x^{\prime}-x)$. The conditions
under which this form of the coupling leads to synchronization have also
been studied \cite{boccaletti}. As can be seen from the examples
presented here, the systems synchronize when the effective coupling exceeds a
threshold strength (Figs.~2(b) and 4(d)). The rate of growth of the phase
difference vanishes at this threshold as a power \cite{wang}.

Stochastic oscillators appear to effectively synchronize independent of the size of fluctuations,  
in a manner similar to chaotic synchronization \cite{pecora}. However,
even when the parameters are identical, exact synchronization is not possible
and the systems can only phase--synchronize.
 
The synchronization  of stochastic oscillators thus has similarities to
the analogous process in deterministic dynamical systems (both with and
without added noise), but also important differences. The mechanisms that we
have described here pertain to systems with intrinsic stochasticity, and are
therefore in a nonperturbative limit.  

The coupling schemes that we have proposed here can find application in the
design and control of synthetic biological networks where synchronous
oscillation may be a desirable feature. McMillen et al.
\cite{kopell} have shown that intercell signaling via a diffusing molecule
can couple genetic oscillators and effect  synchrony.
The present results indicate that such phase synchrony can emerge under very
general conditions with high levels of ambient noise.  In a related vein, one can
speculate that similar mechanisms underlie the synchrony that is so dramatically
evident in cellular processes. As recent time--resolved microarray
experiments of yeast have revealed, the multitude of variable gene
expression patterns classify into a small number of groups, all genes of a
given group having very similar temporal variation profiles \cite{tu}. We have
observed that the above coupling schemes are effective in synchronizing
ensembles of stochastic oscillators \cite{nandi}.   Other physical situations where
such mechanisms  may be relevant are in the study of coupled ecosystems or
coupled weather systems: individual subsystems show stochastic dynamics,
which however have phase synchrony \cite{lloyd}.

Although we have discussed the explicit case of stochastic oscillators,
there is reason to believe that similar microscopic intersystem couplings can bring
about temporal correlations in more general stochastic systems. Investigations of such
phenomena are currently under way \cite{nandi}.

This research is supported by grants from the DBT,
India (to RR) and the CSIR, India through the award of Senior Research
Fellowships (to AN and SG)  and a Research Associateship (to RKBS). We thank M Bennett, T Gross, and J K Bhattacharjee for helpful correspondence.


\begin{thebibliography}{abc99}
\bibitem{lange} In practice, at the end of each integration step, a term
$\gamma\eta X^{1/2}_{i}$ is added to the solution of the deterministic
equations of motion, $\gamma$ being the strength of the Gaussian noise
$\eta$ \cite{cle}.
\bibitem{param} We take the following values for the reaction rates
\cite{genosc}:
$\aa = 50 \h^{-1}$, $\apa = 500 \h^{-1}$, $\ar = 0.01 \h^{-1}$,
$\as = \ar + 0.001 \h^{-1}$, $\apr = 50 \h^{-1}$,
$\aps = \apr + 5 \hr^{-1}$, $\ba = 50 \h^{-1}$, $\br = 5 \h^{-1}$,
$\bs = \br + 0.5 \hr^{-1}$, $\dma = 10 \h^{-1}$, $\dmr = 0.5 \h^{-1}$,
$\dms = \dmr + 0.05 \hr^{-1}$, $\da = 1 \h^{-1}$, $\dr = 0.2 \h^{-1}$,
$\ds = \dr + 0.02 \hr^{-1}$, $\ga = 1 \mol^{-1} \hr^{-1}$,
$\gr = 1 \mol^{-1} \hr^{-1} = \gs$, $\gc = 1 \mol^{-1} \hr^{-1} = \gcp$,
$\ta = 50 \hr^{-1}$, $\tr = 100 \hr^{-1} = \ts$. The initial conditions
are  $\DA=\DR=\DS=1 \mol, \DAP=\DRP=\DSP=\MA=\MR=\MS=A=R=C=0$, $S=1,
C^{\prime}= 1$. The cell has a single copy of the activator and repressor
genes: $\DA + \DAP = \DR + \DRP = \DS + \DSP =1 \mol$, and the volume is
assumed to be unity.
\end{thebibliography}
\end{document}